\documentclass[aps,prd,amsmath,amssymb,superscriptaddress,twocolumn,nofootinbib,showpacs,preprintnumbers,notitlepage]{revtex4-1}

\usepackage{color,paralist}
\usepackage{mathrsfs}
\usepackage{amsfonts,amsmath,amssymb,bm}
\usepackage{epsfig}
\usepackage{graphicx}
\usepackage{amsfonts}
\usepackage{slashed}
\usepackage{float}
\usepackage{cancel}
\usepackage{relsize}
\usepackage[dvipsnames]{xcolor}
\usepackage[colorlinks, citecolor=blue, linktoc=all, linkcolor=blue, urlcolor=blue]{hyperref}
\usepackage[normalem]{ulem}

\newcommand{\nn}{\nonumber}
\newcommand{\xb}{x}

\newcommand{\beq}{\begin{equation}}
\newcommand{\eeq}{\end{equation}}

\newcommand{\ber}{\begin{eqnarray}} 
\newcommand{\eer}{\end{eqnarray}}

\newcommand{\jlab}{Jefferson Lab, Newport News, Virginia 23606, USA}
\newcommand{\adelaide}{\mbox{CSSM and CDMPP, Department of Physics, University of Adelaide, Adelaide 5005, Australia}}

\newcommand{\ut}{Institute for Theoretical Physics, T\"ubingen University, Auf der Morgenstelle 14, 72076 T\"ubingen, Germany}
\newcommand{\ur}{Institute for Theoretical Physics, Regensburg University, 93040 Regensburg, Germany}

\allowdisplaybreaks

\begin{document}

\preprint{JLAB-THY-23-3739, ADP-22-38/T1209}
\title{Resonant contributions to polarized proton structure functions}
\author{A.~N.~\surname{Hiller Blin}}
\affiliation{\ur}
\affiliation{\ut}
\author{V.~I.~\surname{Mokeev}}
\affiliation{\jlab}
\author{W.~\surname{Melnitchouk}}
\affiliation{\jlab}
\affiliation{\adelaide}

\begin{abstract}
Nucleon resonance contributions to the polarized proton $g_1$ and $g_2$ structure functions are computed from resonance electroexcitation amplitudes extracted from CLAS exclusive meson electroproduction data.
Including resonances in the mass range up to 1.75~GeV, and taking into account the interference between excited states, we compare the resonant contributions with the  polarized proton structure function and polarization asymmetry data from Jefferson Lab 6~GeV measurements.
All resonance-like structure observed in the polarized structure functions and asymmetries can be attributed to the resonant contributions, confirming their essential role in the behavior of $g_1$ and $g_2$ in the resonant region over the entire range $Q^2 < 7.5$~GeV$^2$ covered by the measurements.
Comparing the resonance contributions with the $g_1$ and $g_2$ structure functions computed from parton distribution functions extrapolated from the deep-inelastic region, we also quantify the degree to which quark-hadron duality holds for $g_1$ and $g_2$ and their moments.
\end{abstract}

\date{\today}
\maketitle

%%%%%%%%%%%%%%%%%%%%%%%%%%%%%%%%%%%%%%%%%%%%%%%%%%%%%%%%%%%%%%%%%%%%%%%%%%%%%
\section{Introduction}

Inclusive electron scattering from proton targets has traditionally played a key role in the evolution of our understanding of nucleon structure~\cite{Jimenez-Delgado:2013sma, Gao:2017yyd, Ethier:2020way}. 
High-energy cross sections, expressed in terms of structure functions, have driven global QCD analyses of parton distribution functions (PDFs)
%Harland-Lang:2014zoa, Dulat:2015mca, Ball:2017nwa
~\cite{Hou:2019efy,Bailey:2020ooq,NNPDF:2021njg, Moffat:2021dji, Accardi:2016qay, Sato:2019yez, Alekhin:2017kpj, Cocuzza:2022jye} through QCD factorization theorems~\cite{Collins:1989gx}.
Historically, global QCD analyses based on the leading twist approximation have accurately described data for invariant masses of the final state hadrons as low as $W \approx 2$~GeV, just above the region populated by prominent resonances, and photon virtualities $Q^2 \gtrsim 1-2$~GeV$^2$.
Typically, cuts on these variables are made to exclude the resonance region, which cannot be described in terms of perturbatively factorized hard scattering amplitudes and nonperturbative distributions.
An important question, however, is how low in $W$ and $Q^2$ can one go while still retaining a partonic interpretation of the scattering process. 
Some global analyses have included extensions down to $W \approx 1.75$~GeV~\cite{Owens:2012bv, Accardi:2016qay}, which requires careful treatment of subleading effects, such as target mass corrections~\cite{Accardi:2016qay, Alekhin:2017kpj, Moffat:2019qll, Brady:2011uy}, higher twists, and factorization breaking corrections.

In addition to the prospect of providing stronger constraints on PDFs at large parton momentum fractions $x$, the resonance--scaling transition region is important for understanding fundamental emergent features of QCD~\cite{Melnitchouk:2005zr,Lagerquist:2022tml}.
In nature, a duality has been observed between the behavior of structure functions in the nucleon resonance region, averaged over specific energy intervals, and the scaling function in the deep-inelastic scattering (DIS) region at higher $W$~\cite{Bloom:1970xb, Osipenko:2003bu, Prok:2014ltt, Malace:2009kw, Christy:2007ve, Tvaskis:2016uxm, Liang:2004tj}. 
Precision measurements of inclusive electron scattering cross sections at Jefferson Lab in the resonance region underlie these observations in unpolarized proton structure functions~\cite{Niculescu:2000tj, Niculescu:2000tk, Liuti:2001qk, Niculescu:2015wka, Forte:1998nw, Forte:2000wh}. 
The experimental improvements in inclusive scattering of polarized electron beams from polarized nucleon targets have also allowed the extension of duality studies to spin-dependent nucleon structure functions~\cite{RSS:2006tbm, RSS:2008ceg, Bosted:2006gp, CLAS:2006ozz, Dharmawardane:2004yw, SANE:2018pwx}, which, in contrast to the spin-averaged case, are not positive definite quantities.
Duality and scaling in spin-dependent structure functions have also been explored theoretically in Refs.~\cite{Close:1972ah, Carlson:1998gf, Edelmann:1999yp, Melnitchouk:2005zr}.
It was found that, for example, in comparing with global QCD parametrizations of polarized structure functions~\cite{Gluck:1995yr, Gluck:2000dy}, an onset of duality could be seen at $Q^2$ as low as $1-2$~GeV$^2$.

Understanding the functional dependence of duality on $W$ requires a theoretical understanding of how a smooth scaling function can arise from a sum of sharp resonances~\cite{Domokos:1971ds, Domokos:1971bw, Domokos:1972yc, Feynman:1971wr, Rein:1980wg, Isgur:2001bt, Jeschonnek:2002db, Davidovsky:2002nj, Fiore:2002re, Jenkovszky:2012dc, DeRujula:1976baf, Simula:1999rr, Ent:2000jj, Melnitchouk:2001eh, Kulagin:2021mee, Kamano:2013iva, Nakamura:2015rta, Nakamura:2016cnn}.
Central to this was the observation, made by Close and Isgur~\cite{Close:2001ha}, that in certain simplified cases the square of a sum of amplitudes for transitions from ground to excited states can reduce to a sum of squares of amplitudes, as would be appropriate for incoherent scattering from partons in the nucleon.
In more physically realistic scenarios, the realization of this idea was elaborated for the spin-flavor symmetric quark model~\cite{Close:2003wz, Close:2009yj}, where cancellations between even and odd parity multiplets ensure the vanishing of coherent interference contributions. 
On the other hand, while presenting a simple intuitive picture of the connection between resonances and scaling contributions, absent from such discussions are considerations of the role of the nonresonant background, the incorporation of which is rather challenging within a consistent framework that treats resonant and nonresonant contributions on similar footing.

While a quantitative description of the latter from first principles is currently beyond reach, insight may be obtained from phenomenological analyses.
For example, the experimental program exploring exclusive $\pi^+ n$, $\pi^0 p$, $\eta p$, and $\pi^+ \pi^-p$ electroproduction channels in the resonance region with the CLAS detector at Jefferson Lab has provided important new information on the $\gamma^* p N^*$ electrocouplings of most nucleon resonances in the mass range $W \leq 1.75$~GeV and $Q^2 \leq 5$~GeV$^2$~\cite{Aznauryan:2011qj, Aznauryan:2009mx, Mokeev:2012vsa, Mokeev:2015lda, Park:2014yea, Carman:2020qmb, Mokeev:2020hhu}.
These results make it timely to quantitatively evaluate the evolution of the resonant contributions to inclusive electron scattering observables in the resonance region, using parameters of the individual nucleon resonances extracted from data. 
Such studies were performed for spin-averaged observables in our previous work~\cite{Blin:2019fre, Blin:2021twt}, shedding light into the prospects for electrocoupling studies at $Q^2 > 5$~GeV$^2$ with the CLAS12 detector at Jefferson Lab~\cite{Carman:2020qmb,Proceedings:2020fyd}.
In the present work, we extend the previous phenomenological studies to the spin-dependent structure functions, with the aim of providing insights into the spin dependence of quark-hadron duality and the DIS--resonance transition region.

In Sec.~\ref{sec:polSFex} we give a brief review of the basic formulas for the extraction of spin-dependent structure functions from inclusive polarized electron--proton scattering.
The formulas for the resonant contributions to the unpolarized and polarized structure functions are presented in Sec.~\ref{sec:res_eval} in terms of the $\gamma^* p N^*$ electrocouplings for transverse and longitudinal photons.
Results for the resonant contributions to the inclusive $g_1$ and $g_2$ structure functions are given in Sec.~\ref{sec:results}, where we compare the various resonance contributions to the structure function data.
We also quantify the role of interferences in the sum over resonant amplitudes in both the polarized and unpolarized structure functions.
In Sec.~\ref{sec:duality} we consider the spin dependence of quark-hadron duality in the resonance--DIS transition region for the $g_1$ structure function and the lowest moments of $g_1$ and $g_2$ by comparing the resonance contributions with the corresponding functions extrapolated from the high-$W$ region.
Finally, in Sec.~\ref{sec:outlook} we summarize our findings and discuss future extensions of the study and applications of duality in inclusive and exclusive reactions.

%%%%%%%%%%%%%%%%%%%%%%%%%%%%%%%%%%%%%%%%%%%%%%%%%%%%%%%%%%%%%%%%%%%%%%%%%%%%%
\section{Spin structure functions and polarization asymmetries}
\label{sec:polSFex}

In this section we summarize the polarized electron--nucleon scattering formalism, including the definitions of the spin-dependent $g_1$ and $g_2$ structure functions and inclusive electron scattering observables with polarized beam and target.
In evaluating the resonant contributions, it is convenient to express $g_1$ and $g_2$ in terms of the invariant mass $W$ of the virtual photon--target proton (or final state hadron) system, and the photon virtuality $Q^2>0$ (corresponding to the negative of the four-momentum squared of the virtual photon).

The polarized structure functions are obtained from measurements of double beam-target polarization asymmetries, $A_\parallel$ and $A_\perp$, defined as
\begin{eqnarray}
\label{asym_defin}
A_\parallel
= \frac{Y_{\uparrow \downarrow} - Y_{\uparrow \uparrow}}{Y_{\uparrow \downarrow} + Y_{\uparrow \uparrow}}, 
\qquad
A_\perp
= \frac{Y_{\uparrow \rightarrow} - Y_{\uparrow \leftarrow}}{Y_{\uparrow \rightarrow} + Y_{\uparrow \leftarrow}},
\end{eqnarray}
where $Y_{\uparrow \uparrow}$ and $Y_{\uparrow \downarrow}$ are the scattered electron yields in a $(W,Q^2)$ bin measured for parallel and antiparallel orientations of the electron beam and proton target polarization vectors, respectively, while $Y_{\uparrow \rightarrow}$ and $Y_{\uparrow \leftarrow}$ are the corresponding yields for longitudinally polarized electrons for two opposite orientations of the proton polarization vector transverse to the laboratory floor plane.
This plane is irrelevant in computations of reaction amplitudes, but it is the only plane relative to the proton polarization vector that can be oriented in the measurements. 
In the one-photon exchange approximation, $A_\parallel$ and $A_\perp$ should neither be affected by the uncertainties of the scattered electron detection efficiency, nor by the uncertainties of the overall normalization, as such uncertainties cancel between the numerator and denominator.

The measured asymmetries $A_\parallel$ and $A_\perp$ are related to the virtual photon asymmetries $A_1$ and $A_2$ according to~\cite{Anselmino:1994gn, SANE:2018pwx, roberts_1990}
\begin{subequations}
\label{eq:A12}
\begin{eqnarray}
    A_1
    &=&\frac{1}{(E+E^\prime)D^\prime}
    \left[ (E-E^\prime\cos\theta) A_\parallel
         - \frac{E^\prime\sin\theta}{\cos\phi} A_\perp
    \right],
\nonumber\\
& &
\\
    A_2
    &=&\frac{\sqrt{Q^2}}{2ED^\prime}
    \left[ A_\parallel 
         + \frac{E-E^\prime\cos\theta}{E^\prime\sin\theta\cos\phi} A_\perp
    \right],
\end{eqnarray}
\end{subequations}
where $E$ and $E'$ are the energies of the incoming and scattered electron, respectively. 
The polar angle of the scattered electron is $\theta$, while $\phi$ is the azimuthal angle for a proton polarization in the lab frame with $z$-axis along the initial electron momentum and $x$ ($y$)-axis normal (in) to the scattering plane.
The factor $D^{\prime}$ is given by
\begin{eqnarray}
\label{Dpar}
    D^\prime &=& \frac{1-\epsilon}{1+\epsilon R},
\end{eqnarray} 
where 
\begin{eqnarray}
    \epsilon &=& \left(1+2\,\frac{Q^2+\nu^2}{Q^2}\tan^2\frac{\theta}{2}\right)^{-1}
\end{eqnarray}
is the virtual photon polarization~\cite{Tiator:2011pw}, and $R$ represents the ratio of longitudinally ($\sigma_L$) to transversely ($\sigma_T$) polarized virtual photon absorptive cross sections, $R=\sigma_L/\sigma_T$.
The energy transfer $\nu$ to the proton target in the lab frame is related to the invariant mass $W$ of the virtual photon--target proton system,
$\nu = (W^2-M^2+Q^2)/2M = Q^2/2Mx$,
where $x=Q^2/2M\nu$ is the Bjorken scaling variable and $M$ is the nucleon mass.

The cross sections for virtual photons in electron scattering are related to hadron electroproduction cross sections through the virtual photon flux, determined by the electron scattering kinematics~\cite{Blin:2021twt, Blin:2019fre}.
In inclusive scattering of unpolarized electrons from unpolarized protons, only virtual photons of transverse and longitudinal polarizations contribute to the absorptive cross sections, $\sigma_T$ and $\sigma_L$, respectively.
For a proton polarization vector aligned along the virtual photon three-momentum, the transverse cross section is determined by the sum of helicity-1/2 and 3/2 contributions, 
\begin{align}
\label{sigt}
\sigma_T = \frac12 \Big( \sigma_T^{1/2} + \sigma_T^{3/2} \Big),    
\end{align}
where the superscripts 1/2 and 3/2 refer to the absolute value of the sum of the spin projections of the virtual photon and proton in the direction of the virtual photon three-momentum~\cite{roberts_1990}.
For a proton polarization vector with normal component relative to the photon direction, the virtual photon cross section receives additional interference contributions.   
In terms of cross sections, the virtual photon asymmetries are given by~\cite{roberts_1990, Dharmawardane:2004yw, Melnitchouk:2005zr}
\begin{align}
\label{a1a2} 
A_1 &= \frac{\sigma_T^{1/2}-\sigma_T^{3/2}}{2\sigma_T}, 
\qquad A_2 = \frac{\sigma_{I}}{\sigma_T},
\end{align}
where $\sigma_I$ is the real part of the interference amplitude for virtual photons with longitudinal and transverse polarizations. 
The $A_1$ and $A_2$ can be extracted from the measured $A_\parallel$ and $A_\perp$ asymmetries via Eqs.~(\ref{eq:A12}).

The polarized structure functions $g_1$ and $g_2$ are expressed in terms of the $A_1$ and $A_2$ virtual photon asymmetries as
\begin{subequations}
\label{eq:g1g2} 
\begin{align}
%g_1 &= \frac{\nu^2}{\nu^2+Q^2}\,F_1 \bigg( \frac{\sqrt{Q^2}}{\nu}A_2+A_1 \bigg),
g_1 &= \frac{1}{\rho^2}\, F_1 
        \Big( A_1 + A_2\sqrt{\rho^2-1} \Big),
\\
%g_2 &= \frac{\nu^2}{\nu^2+Q^2}\,F_1 \bigg( \frac{\nu}{\sqrt{Q^2}}A_2-A_1 \bigg).
g_2 &= \frac{1}{\rho^2}\, F_1 
        \Big( -A_1 + \frac{A_2}{\sqrt{\rho^2-1}} \Big),
\end{align}
\end{subequations}
where the kinematic factor $\rho^2 = 1 + Q^2/\nu^2$, and the unpolarized structure functions $F_1$ and (for completeness) $F_2$ are given by~\cite{Drechsel:2002ar, Melnitchouk:2005zr},
\begin{subequations}
\begin{eqnarray}
\label{eq:F1} 
F_1 &=& \frac{{K M}}{4\pi^2\alpha}\, \sigma_T,
\\
F_2 &=& \frac{{K \nu}}{4\pi^2\alpha}\frac{{ Q^2}}{\nu^2+Q^2}[\sigma_T+\sigma_L].
\end{eqnarray}
\end{subequations}
Here, $\alpha$ is the electromagnetic fine structure constant, and $K=(W^2-M^2)/2M$ is the equivalent photon energy in the Hand convention.
Finally, we can also define the individual helicity structure functions $H_{1/2}$ and $H_{3/2}$ for helicity-1/2 and 3/2 contributions as~\cite{Malace:2011ad}
\begin{subequations}
\label{E:h12_h32}
\begin{align}
%H_{\frac12} &= F_1 + g_1 - \frac{Q^2}{\nu^2}g_2, \\
%H_{\frac32} &= F_1 - g_1 + \frac{Q^2}{\nu^2}g_2,
H_{1/2} &= F_1 + g_1 - (\rho^2-1)\, g_2,
\\
H_{3/2} &= F_1 - g_1 + (\rho^2-1)\, g_2. 
\end{align}
\end{subequations}

For the polarized structure function measurements in Jefferson Lab Hall~B, because of the CLAS detector's nearly 4$\pi$ acceptance, the measured $g_1$ structure function was obtained over the range of $W$ up to $\approx 1.8$~GeV, in any given $Q^2$ bin~\cite{CLAS:2006ozz, Dharmawardane:2004yw}. 
However, since only longitudinally polarized proton targets have been used in CLAS electroproduction experiments, the $g_1$ structure function was inferred from the CLAS data by employing a fit to the $g_2$ world data, as described in Ref.~\cite{CLAS:2017qga}.
In contrast, for measurements of inclusive beam-target polarization asymmetries with  spectrometers of small acceptance, such as in Jefferson Lab Hall~C~\cite{RSS:2006tbm, RSS:2008ceg, SANE:2018pwx, JeffersonLabHallAg2p:2022qap}, both $A_\parallel$ and $A_\perp$ have been determined, allowing the full reconstruction of $g_1$ and $g_2$ from the measured observables. 
On the other hand, because of the limited detector acceptance the polarized structure functions at a given $Q^2$ value there were determined for a narrow $W$ range only. \\

%%%%%%%%%%%%%%%%%%%%%%%%%%%%%%%%%%%%%%%%%%%%%%%%%%%%%%%%%%%%%%%%%%%%%%%%%%%%%
\section{Resonant contributions to inclusive structure functions}
\label{sec:res_eval}

In this section we describe the evaluation of the resonant contributions to the $g_1$ and $g_2$ structure functions, extending the formalism used in the analysis of unpolarized structure functions in Refs.~\cite{Blin:2019fre, Blin:2021twt} to the polarized sector. 
To take into account the interference between different resonances, each contribution to the structure functions is evaluated in terms of a coherent sum of resonance amplitudes.
The contribution from each resonance of spin $J$, isospin $I$, and parity $\eta$ can be described by the amplitudes $G_{m}^R$, where $m = +1, 0, -1$ represents the virtual photon spin projection onto the $z$-axis, aligned along the direction of the virtual photon momentum.
Adding the amplitudes coherently, the sum of contributions from the resonances $R$ to the inclusive structure functions can be written as~\cite{Carlson:1998gf, Melnitchouk:2005zr}
\begin{widetext}
\begin{subequations}
\label{Eq:coherent}
\begin{eqnarray}
   \left(1+\frac{Q^2}{\nu^2}\right) g_1^\text{res}
%   \rho^2\, g_1^R
   &=& M^2\sum_{IJ\eta}
   \Bigg\{
     \bigg| \sum_{R^{IJ\eta}} G_+^{R^{IJ\eta}} \bigg|^2
   - \bigg| \sum_{R^{IJ\eta}} G_-^{R^{IJ\eta}} \bigg|^2
\nn\\
    && \qquad\qquad +\, 
    \frac{\sqrt{2Q^2}}{\nu}\,
%    \sqrt{2(\rho^2-1)}\,
    \Re\bigg[
    \bigg( \sum_{R^{IJ\eta}} G_0^{{R^{IJ\eta}}} \bigg)
    \bigg( \sum_{R^{IJ\eta}}(-1)^{J_{R^{IJ\eta}} - \frac12}\, \eta_{R^{IJ\eta}}
    G_+^{R^{IJ\eta}} \bigg)^\ast
    \bigg]
    \Bigg\},
\\
    \left(1+\frac{Q^2}{\nu^2}\right) g_2^\text{res}
%    \rho^2\, g_2^R
    &=& -M^2 \sum_{IJ\eta}
    \Bigg\{
      \bigg| \sum_{R^{IJ\eta}} G_+^{R^{IJ\eta}} \bigg|^2 
    - \bigg| \sum_{R^{IJ\eta}} G_-^{R^{IJ\eta}} \bigg|^2
\nn\\
    && \qquad\qquad -\, 
    \frac{\nu\sqrt{2}}{\sqrt{Q^2}}\,
%    \frac{\sqrt2}{\sqrt{\rho^2-1}}\,
    \Re\bigg[
    \bigg( \sum_{R^{IJ\eta}} G_0^{{R^{IJ\eta}}} \bigg) 
    \bigg( \sum_{R^{IJ\eta}}(-1)^{J_{R^{IJ\eta}}-\frac12}\, \eta_{R^{IJ\eta}}
    G_+^{R^{IJ\eta}} \bigg)^\ast
    \bigg]
    \Bigg\},
\end{eqnarray}
for the spin-dependent structure functions, and
\begin{eqnarray}
    F_1^\text{res}
    &=& M\sum_{IJ\eta}
    \Bigg\{
      \bigg| \sum_{R^{IJ\eta}} G_+^{R^{IJ\eta}} \bigg|^2
    + \bigg| \sum_{R^{IJ\eta}} G_-^{R^{IJ\eta}} \bigg|^2
    \Bigg\},
\\
    \left(1+\frac{\nu^2}{Q^2}\right) F_2^\text{res}
%    \frac{\rho^2}{Q\sqrt{\rho^2-1}}\, F_2^R
    &=& M\nu\sum_{IJ\eta}
%    &=& M\sum_{IJ\eta}
    \Bigg\{
      \bigg|  \sum_{R^{IJ\eta}} G_+^{R^{IJ\eta}} \bigg|^2
    + \bigg|  \sum_{R^{IJ\eta}} G_-^{R^{IJ\eta}} \bigg|^2
    + 2\bigg| \sum_{R^{IJ\eta}} G_0^{R^{IJ\eta}} \bigg|^2
    \Bigg\},
\label{Eq:coherentF2}
\end{eqnarray}
\end{subequations}
\end{widetext}
for the spin-averaged structure functions.
The outer sums in Eqs.~(\ref{Eq:coherent}) run over the possible values of the spin $J$, isospin $I$, and intrinsic parity $\eta$, while the inner sums run over all resonances $R^{IJ\eta}$ that satisfy $J_R=J$, $I_R=I$ and $\eta_R=\eta$ for the spin, isospin and parity of the resonance $R$~\footnote{Note that in Eq.~(12b) of Ref.~\cite{Blin:2021twt} the overall factor should be $1+\nu^2/Q^2$, as in our Eq.~(\ref{Eq:coherentF2}), instead of $1+Q^2/\nu^2$.}.

The amplitudes $G_{m}^R$ in Eqs.~(\ref{Eq:coherent}) are related to the resonance electrocouplings $A_{1/2}^R$, $A_{3/2}^R$, and $S_{1/2}^R$~\cite{Aznauryan:2011qj, Blin:2021twt} according to
\begin{subequations}
\label{gm_amplitudes}
\begin{align}
G_+^R &= C \frac{\sqrt{M_R\Gamma_R(W)}}{M_R^2-W^2-i\Gamma_R(W)M_R}\, A^R_{1/2}(Q^2),
\\
G_-^R &= C \frac{\sqrt{M_R\Gamma_R(W)}}{M_R^2-W^2-i\Gamma_R(W)M_R}\,P\, A^R_{3/2}(Q^2),
\\
G_0^R &= C \frac{\sqrt{M_R\Gamma_R(W)}}{M_R^2-W^2-i\Gamma_R(W)M_R}\,P\, S^R_{1/2}(Q^2),
\end{align}
\end{subequations}
where the coefficient $C$ is given by~\footnote{Note that the coefficient $C$ in Eq.~(16) of Ref.~\cite{Blin:2021twt} should have a factor $2\pi$ in the denominator instead of $4\pi$.}
\begin{align}
   C &= \frac{1}{2\pi} \sqrt{\frac{W^2-M^2}{\alpha M}} 
   \frac{q_{\gamma, R}}{q_\gamma},
\end{align}
and $P = \eta\, (-1)^{J-1/2}$ is the parity transformation factor, with $M_R$ and $\Gamma_R(W)$ the mass and energy-dependent total hadronic decay width, respectively, for each resonance~$R$.
Details of the computation of the $W$ evolution of $\Gamma_R(W)$ can be found in Ref.~\cite{Blin:2019fre}.
The photon three-momentum $q_\gamma$ in the virtual photon--target proton center of mass frame is given by
\begin{align}
    q_\gamma &= \sqrt{Q^2+E^2_\gamma},
\end{align}
with $q_\gamma^R\equiv q_\gamma(W\!=\!M_R)$ and 
\begin{align}
    E_\gamma &= \frac{W^2-Q^2-M^2}{2W}
\end{align}
is the photon energy.
As detailed in Refs.~\cite{Blin:2019fre, Blin:2021twt}, for the resonance electrocouplings we use the interpolation functions fitted to the results on their $Q^2$ evolution determined from analyses of exclusive meson electroproduction channels in the resonance region~\cite{Mokeev:2022xfo, Carman:2020qmb} (see Refs.~\cite{CLAS:coups, CLAS:coupsDB}). \\

%%%%%%%%%%%%%%%%%%%%%%%%%%%%%%%%%%%%%%%%%%%%%%%%%%%%%%%%%%%%%%%%%%%%%%%%%%%%%
\section{Comparison with polarized inclusive observables}
\label{sec:results}

Having outlined the relevant formulas needed to compute the resonant contributions to the polarized structure functions from resonance electrocouplings~\cite{Mokeev:2022xfo, Carman:2020qmb}, in this section we present the numerical results and compare them with double beam--target spin asymmetries measured at Jefferson Lab~\cite{Dharmawardane:2004yw, CLAS:2006ozz, CLAS:2021apd, RSS:2006tbm, RSS:2008ceg, SANE:2018pwx}.
In practice, studies of the resonant contributions are limited to the range $W \lesssim 1.8$~GeV and $Q^2 \lesssim 5$~GeV$^2$, where the resonance electrocouplings are currently available \cite{Mokeev:2022xfo, Carman:2020qmb, Blin:2019fre}. 
The list of nucleon resonances included in the computations of the resonant contributions to spin observables in this work is as in the previous study of the unpolarized $F_1$ and $F_2$ structure functions in Ref.~\cite{Blin:2019fre}.
% CLAS~\cite{Dharmawardane:2004yw, CLAS:2006ozz, CLAS:2021apd}
% RSS~\cite{RSS:2006tbm, RSS:2008ceg}
% SANE~\cite{SANE:2018pwx}

%............................................................................
\subsection{Resonant contributions to $g_1$ and $g_2$}

\begin{figure*}[th]
\includegraphics[width=0.8\textwidth]{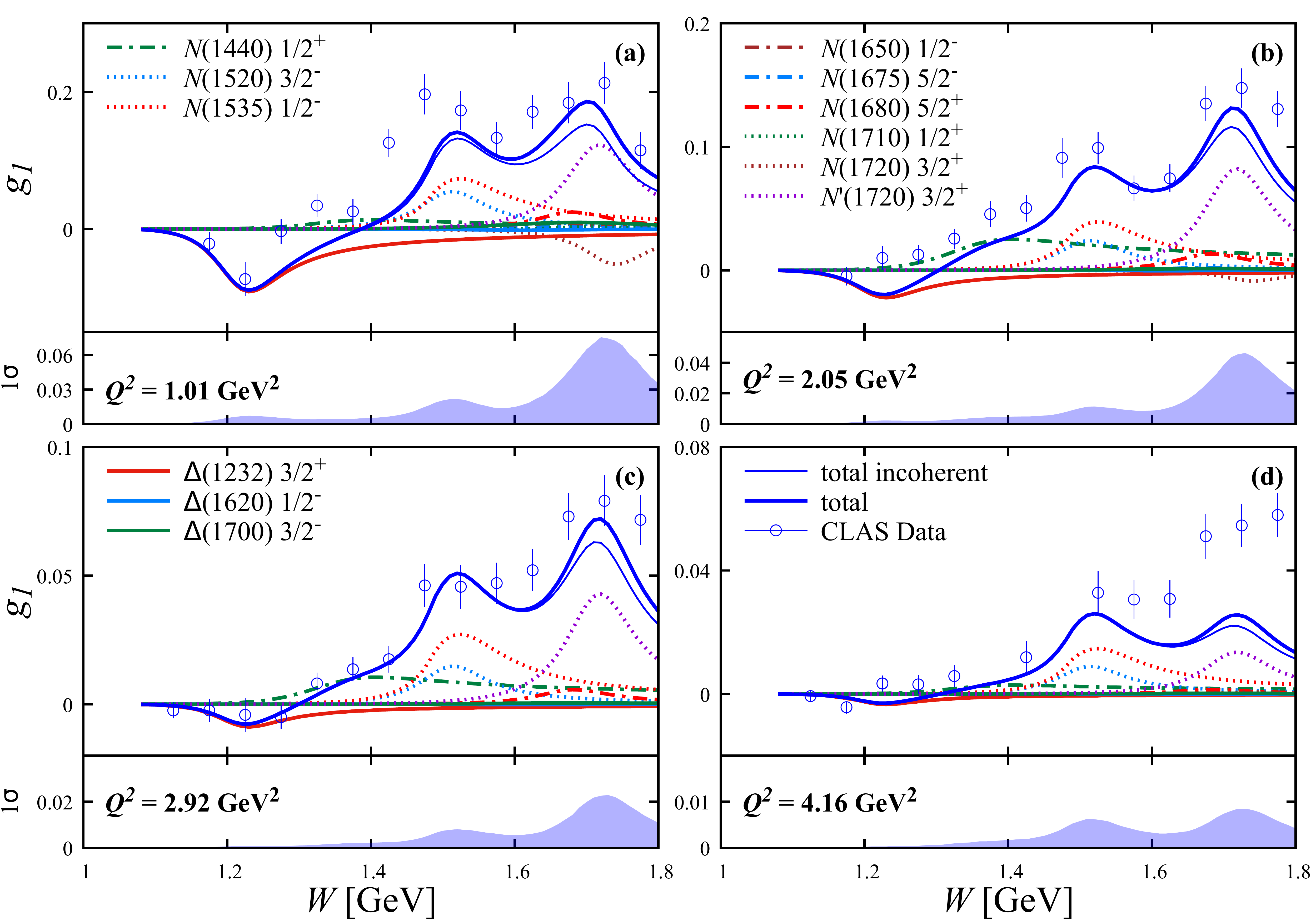}
\caption{Proton $g_1$ structure function versus $W$ at fixed $Q^2$ corresponding to the data bins in Refs.~\cite{CLAS:2006ozz, Dharmawardane:2004yw} (open blue circles): 
{\bf (a)}~$Q^2=1.01$~GeV$^2$, 
{\bf (b)}~$Q^2=2.05$~GeV$^2$, 
{\bf (c)}~$Q^2=2.92$~GeV$^2$, 
{\bf (d)}~$Q^2=4.16$~GeV$^2$, compared with resonant contributions computed from the individual $N$ and $\Delta$ states with (thick blue curves) and without (thin blue curves) accounting for the interference between resonances. Individual excited $N$ and $\Delta$ resonance contributions are shown separately. Below each panel the $1\sigma$ uncertainties, computed by propagating electrocoupling uncertainties via a bootstrap approach, on the coherent resonance sum are shown~\cite{Blin:2019fre}.}
\label{F:g1sing}
\end{figure*}

\begin{figure*}[th]
\includegraphics[width=0.8\textwidth]{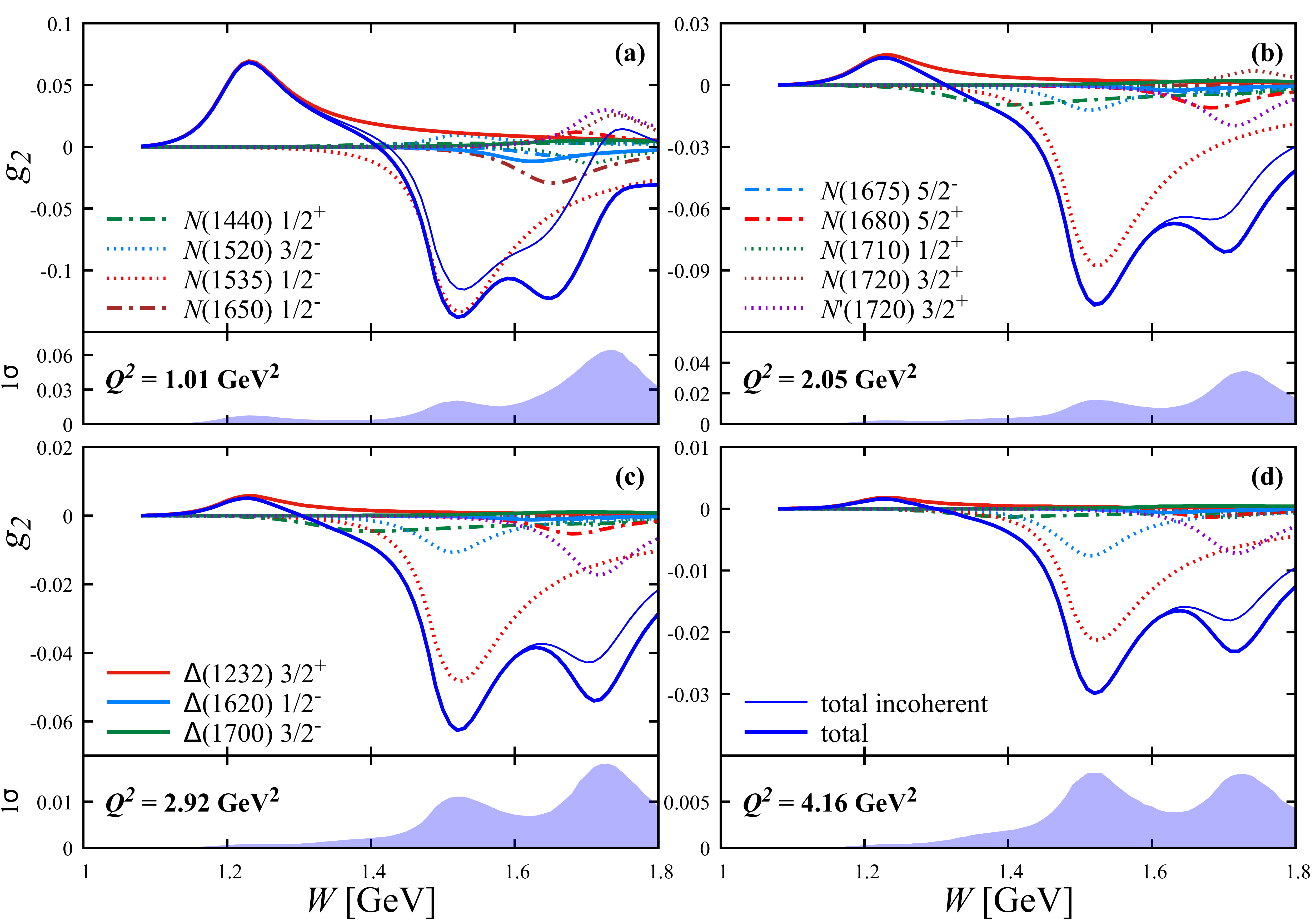}
\caption{Proton $g_2$ structure function versus $W$ at fixed $Q^2$:
{\bf (a)}~$Q^2=1.01$~GeV$^2$, 
{\bf (b)}~$Q^2=2.05$~GeV$^2$, 
{\bf (c)}~$Q^2=2.92$~GeV$^2$, 
{\bf (d)}~$Q^2=4.16$~GeV$^2$. The curves are as in Fig.~\ref{F:g1sing}.}
\label{F:g2sing}
\end{figure*}

The resonance contributions to the proton $g_1$ and $g_2$ structure functions evaluated within the formalism described in Sec.~\ref{sec:res_eval} are shown in Figs.~\ref{F:g1sing} and \ref{F:g2sing}, respectively, as a function of $W$, for several fixed values of $Q^2$ between $Q^2 \approx 1$ and 4~GeV$^2$.
Shown are the individual resonance contributions, as well as the sum of resonances, both with and without accounting for the interference between resonances.
For $g_1$, the calculated structure function is compared with data from CLAS measurements~\cite{CLAS:2006ozz, Dharmawardane:2004yw}.

The impact of the resonance contributions is clearly seen in the evolution of $g_1$ and $g_2$ with $W$ and $Q^2$. 
The dips and peaks in the $W$ dependence in the first resonance region seen in Figs.~\ref{F:g1sing} and \ref{F:g2sing} are directly related to the behavior of the $A_{1/2}$ and $A_{3/2}$ electrocouplings for the $\Delta(1232)\,3/2^+$ resonance, with the absolute values of $A_{3/2}$ remaining larger than that of $A_{1/2}$ over the entire range of $Q^2 \lesssim 5$~GeV$^2$ covered in the analysis.
Consequently, the contribution from the $\Delta(1232)\,3/2^+$ becomes negative for $g_1$ and positive for $g_2$, according to Eqs.~(\ref{Eq:coherent}) and (\ref{gm_amplitudes}). 
The difference between the absolute values of the $A_{1/2}$ and $A_{3/2}$ electrocouplings for the $\Delta(1232)\,3/2^+$ decreases with $Q^2$, resulting in a decrease of the oscillations of the dips and peaks seen in the $W$ dependence of $g_1$ and $g_2$ at larger $Q^2$.

For the states in the second and third resonance regions, apart from the $N(1720)\,3/2^+$, the magnitudes of the $A_{1/2}$ electrocouplings are larger than those of the $A_{3/2}$ electrocouplings \cite{Blin:2019fre}.
The total resonant contributions to both $g_1$ and $g_2$ therefore change sign at $W$ values between the first and second resonance regions.
The sign flip in the $W$ dependence of $g_1$ has been observed in data from CLAS~\cite{CLAS:2006ozz, Dharmawardane:2004yw} (see Fig.~\ref{F:g1sing}), and our analysis confirms that it is indeed driven by the resonant contributions.

Assuming the validity of the Burkhardt-Cottingham sum rule \cite{Burkhardt:1970ti} further imposes a sign flip in the $W$ dependence of the $g_2$ structure function, consistent with the predicted sign flip illustrated in Fig.~\ref{F:g2sing}.
When information about its $W$ dependence in the range of $W \lesssim 1.8$~GeV becomes available from future experiments in any given bin of $Q^2$, it will be interesting to analyse whether the predicted sign flip is driven by the resonances or by the nonresonant contributions at higher~$W$ and $Q^2$.

The peaks in the $g_1$ and $g_2$ structure functions seen in Figs.~\ref{F:g1sing} and \ref{F:g2sing} in the second resonance region arise from the $N(1535)\,1/2^-$ and $N(1520)\,3/2^-$ resonances.
The $W$ dependence of $g_1$ from the experimental data reveals pronounced peaks in the second resonance region in all $Q^2$ bins, suggesting that these resonances are essential in shaping the structure function in this region. 
At the same time, there is a nonvanishing difference between the $W$ dependence of the measured $g_1$ structure function and the computed resonant contributions, especially at lower~$Q^2$. 
The $g_1$ structure function in the second resonance region is therefore determined by both the contributions from the nucleon excited states and by other processes in the virtual photon--target proton interaction. 
The nonresonant contributions to $g_1$ could provide insights into spin-dependent nucleon PDFs at large values of $x \sim 1$ in the context of quark-hadron duality~\cite{Melnitchouk:2005zr}.

In the third resonance region, $1.65 < W < 1.75$~GeV, the differences between the measured $g_1$ structure function and computed resonant contributions at high $Q^2$ are even more pronounced than at lower $W$ values. 
As illustrated in Fig.~\ref{F:g1sing}, for $g_1$ the third resonance peak is mostly driven by the new baryon state $N'(1720)\,3/2^+$~\cite{Mokeev:2020hhu} over the entire $Q^2$ range studied, a finding which is supported by the behavior of the $g_1$ data.
In contrast, the resonant contributions to $g_2$, shown in Fig.~\ref{F:g2sing}, are less sensitive to the impact of the $N'(1720)\,3/2^+$ state, whose contribution is comparable to that from the tail of the $N(1535)\,1/2^-$ in the second resonance region. 
These observations emphasize the importance of accounting for all prominent nucleon resonances in a realistic evaluation of the resonant contributions to the polarized $g_1$ and $g_2$ structure functions.

Note that the double beam--target polarization asymmetries measured with CLAS at Jefferson Lab Hall~B in inclusive electron scattering are available only for a longitudinally polarized proton target~\cite{CLAS:2006ozz, Dharmawardane:2004yw}.
Consequently, to determine the $g_1$ structure function a model for $g_2$ needs to be used.
On the other hand, the asymmetries measured in Jefferson Lab Hall~C with small angular acceptance spectrometers in the SANE~\cite{SANE:2018pwx} and RSS~\cite{RSS:2006tbm, RSS:2008ceg} experiments have used both longitudinally and transversely polarized proton targets.
This allows both $A_1$ and $A_2$, or equivalently $g_1$ and $g_2$, to be extracted from the experimental data without model assumptions.

\begin{figure*}[th]
\includegraphics[width=0.8\textwidth]{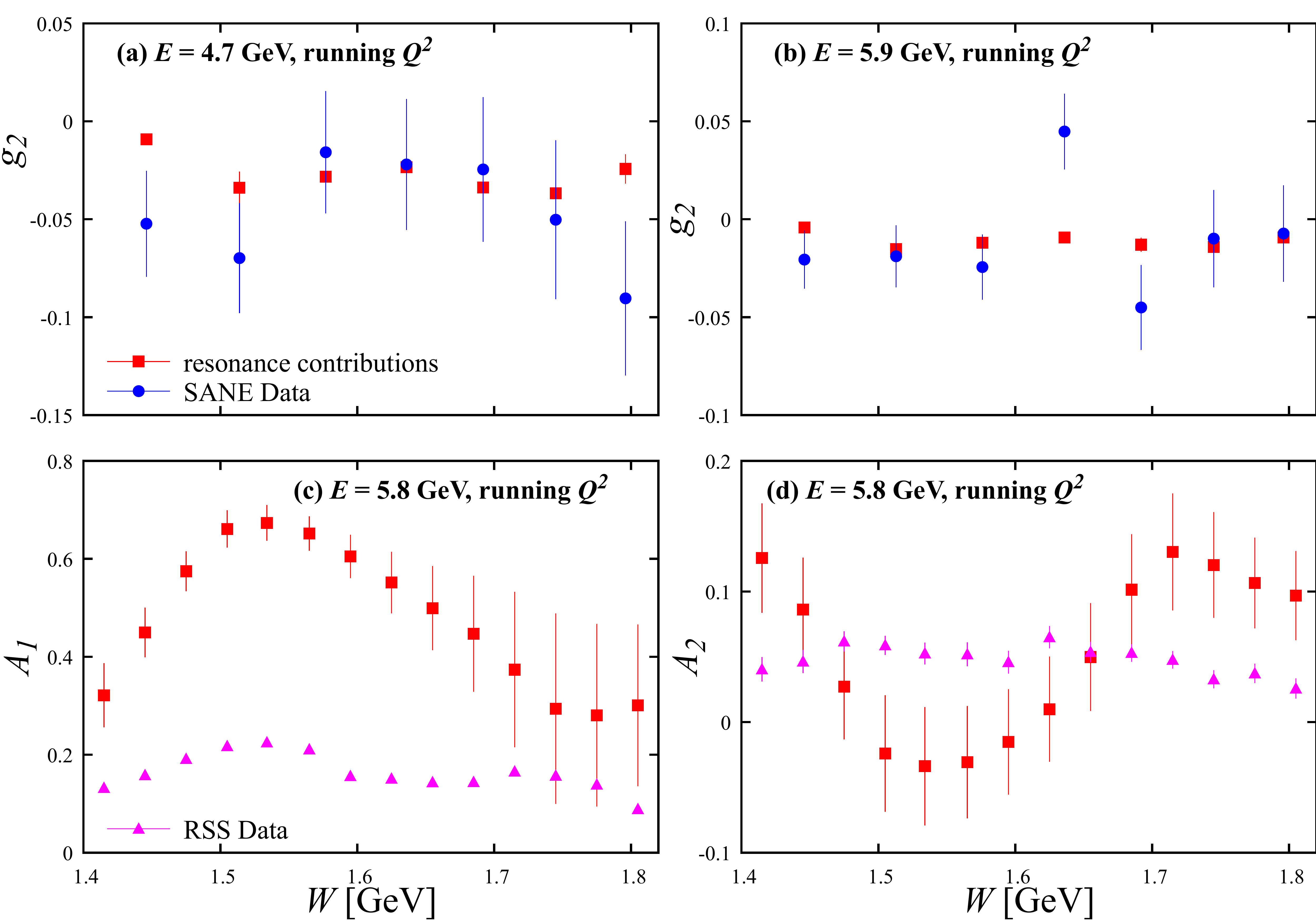}
\caption{Proton $g_2$ structure function (top row) and $A_{1,2}$ asymmetries (bottom row) at running values of $Q^2$ and $W$, corresponding to the data bins in Refs.~\cite{SANE:2018pwx} (blue circles) and \cite{RSS:2006tbm,RSS:2008ceg} (magenta triangles): 
{\bf (a)}~SANE $g_2$ data at $E=4.7$~GeV and $Q^2$ running between $\approx 3.4$ and 4.0~GeV$^2$, 
{\bf (b)}~SANE $g_2$ data at $E=5.9$~GeV and $Q^2$ between $\approx 5.0$ and 5.7~GeV$^2$, 
{\bf (c)}~RSS $A_1$ data at $E=5.8$~GeV and $Q^2$ between $\approx 1.2$ and 1.4~GeV$^2$, 
{\bf (d)}~RSS $A_2$ data at $E=5.8$~GeV and $Q^2$ between $\approx 1.2$ and 1.4~GeV$^2$, compared with the computed total coherent sum of resonant contributions (red squares) to $g_2$, $A_1$, and $A_2$. The uncertainties of the resonant contributions are computed by propagating the electrocoupling uncertainties via a bootstrap approach (see Ref.~\cite{Blin:2019fre} for details).}
\label{F:sane_and_rss}
\end{figure*}

In Fig.~\ref{F:sane_and_rss} we show the computed resonance contributions to $g_2$ compared with the experimental results available from the SANE experiment~\cite{SANE:2018pwx}, as well as the resonant contributions to $A_1$ and $A_2$ compared with the data from the RSS experiment~\cite{RSS:2006tbm, RSS:2008ceg}, in the same $Q^2$ and $W$ bins as where the data were taken.
Unlike with CLAS, because of the small detector acceptance the Hall~C data were taken at running values of both $W$ and $Q^2$.
This has the effect of washing out some of the resonance structure in the $W$ dependence, which is more clearly seen in the high-acceptance CLAS results for $g_1$ [Fig.~\ref{F:g1sing}], and predicted in our computation of the resonance contributions to $g_2$ [Fig.~\ref{F:g2sing}].
This is the case even though the kinematic range is compatible with that of the CLAS data.
This observation underlines the importance of high-acceptance measurements in the resonance region to obtain information on the inclusive structure functions within a broad range of $W$ in any given bin of $Q^2$. 
The comparison between the computed resonant contribution to $g_2$ and the experimental results shown in Fig.~\ref{F:sane_and_rss} suggests that, within the resonance region of $W < 1.75$~GeV, the largest contribution to $g_2$ stems from the resonant part.

%............................................................................
\subsection{Helicity structure functions}

\begin{figure*}[p]
\includegraphics[width=0.8\textwidth]{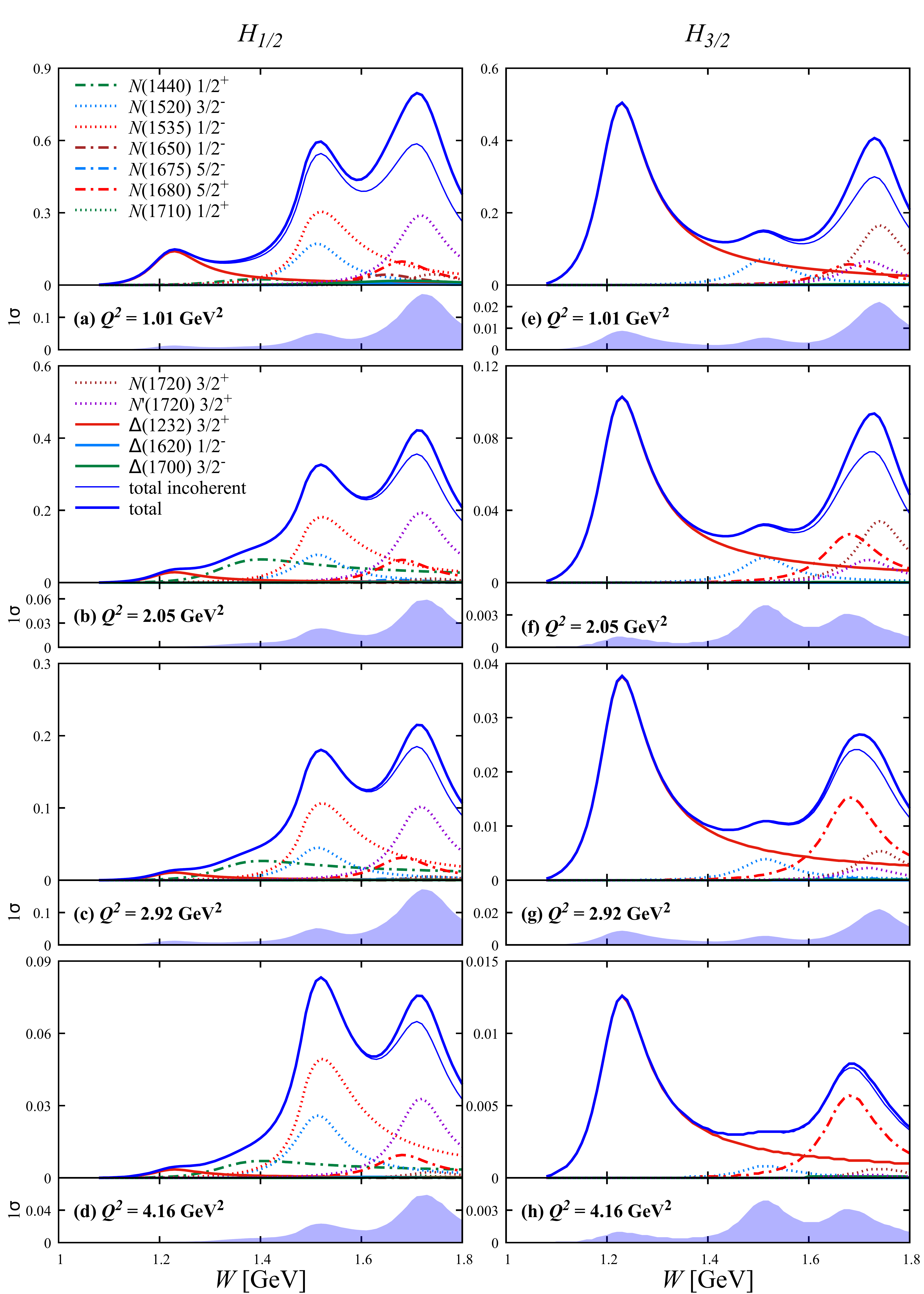}
\caption{Proton helicity structure functions $H_{1/2}$ [{\bf (a)}--{\bf (d)}] and $H_{3/2}$ [{\bf (e)}--{\bf (h)}] at fixed values of $Q^2$ between $Q^2 \approx 1$~GeV$^2$ and 4~GeV$^2$. 
% {\bf (a)} and {\bf (e)}~$Q^2=1.01$~GeV$^2$, 
% {\bf (b)} and {\bf (f)}~$Q^2=2.05$~GeV$^2$, 
% {\bf (c)} and {\bf (g)}~$Q^2=2.92$~GeV$^2$, 
% {\bf (d)} and {\bf (h)}~$Q^2=4.16$~GeV$^2$,
The curves are as described in Fig.~\ref{F:g1sing}.}
\label{F:h1232sing}
\end{figure*}

In Fig.~\ref{F:h1232sing} we show the resonance contributions to the helicity structure functions $H_{1/2}$ and $H_{3/2}$, constructed from the unpolarized $F_1$ and polarized $g_1$ and $g_2$ structure functions.
As for the unpolarized and polarized structure functions, the resonance peaks are also clearly visible for the individual helicity functions.
As expected, the resonant peak { in the first resonance region}, saturated by the $\Delta(1232)\,3/2^+$ resonance, is dominant in $H_{3/2}$, but is vanishingly small for $H_{1/2}$, especially at larger $Q^2$.
The opposite is observed for the peaks in the second resonance region, which are dominated by the contributions from the $N(1520)\,3/2^-$ and $N(1535)\,1/2^-$ states.
A peak in the second resonance region is clearly seen in the resonant contribution to $H_{1/2}$ at $Q^2 \lesssim 5$~GeV$^2$, whilst being barely visible in $H_{3/2}$.
In the third resonance region, several overlapping resonances are relevant: the resonance $N'(1720)\,3/2^+$ \cite{Mokeev:2020hhu} remains the largest contributor to $H_{1/2}$, while in the case of $H_{3/2}$ there is an evolution from a substantial contribution from $N'(1720)\,3/2^+$ at $Q^2 < 2$~GeV$^2$ to the dominant contribution from the spin-5/2 state $N(1680)\,5/2^+$ at $Q^2 > 4$~GeV$^2$.

\begin{figure*}[th]
\includegraphics[width=0.8\textwidth]{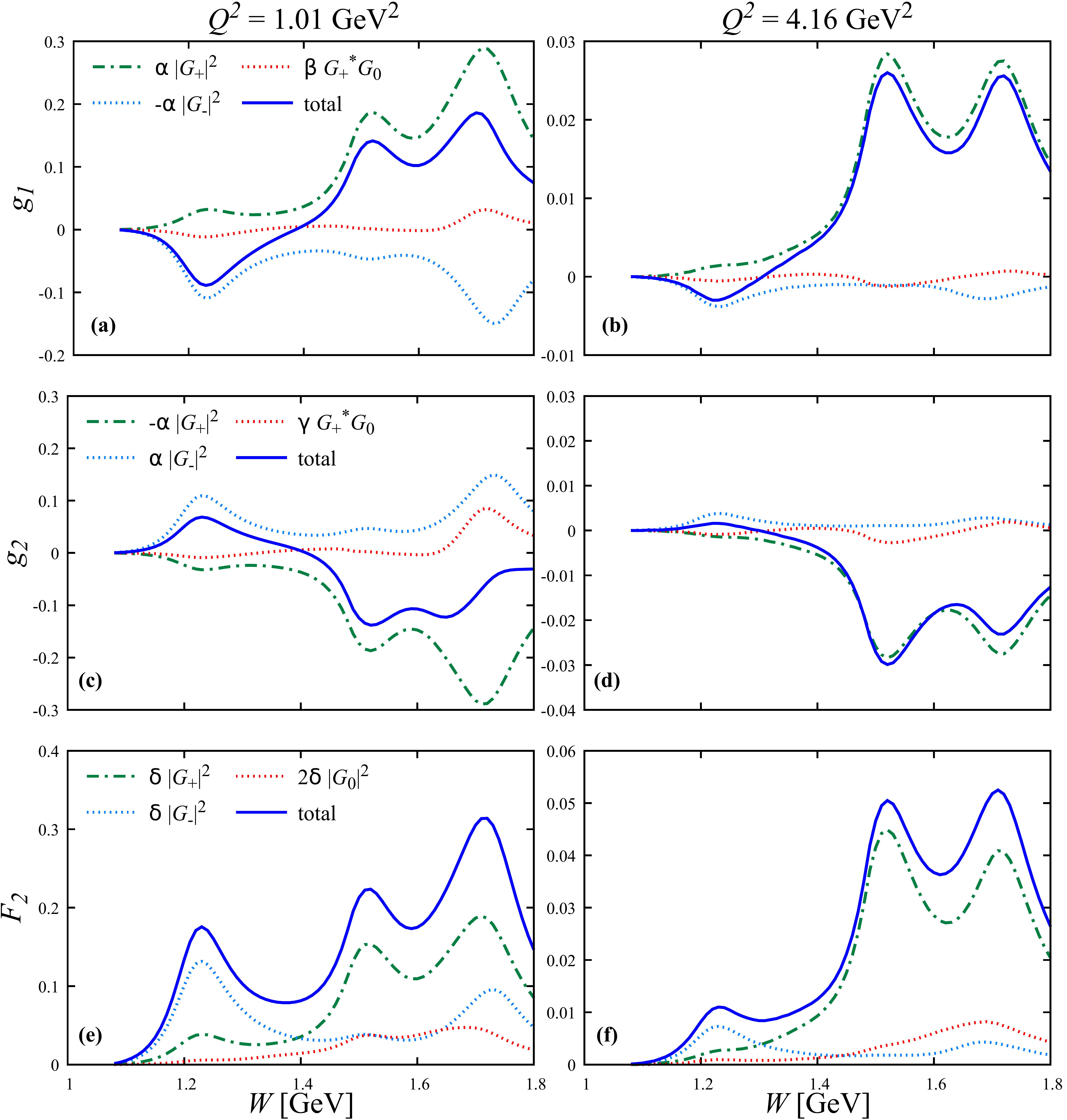}
\caption{Comparison between the total resonant contributions with interferences to the proton $g_1$, $g_2$, and $F_2$ structure functions and the decomposition into $|G_+|^2$, $|G_-|^2$, $|G_0|^2$, and $G_+^*G_0$ at fixed values of $Q^2$:
{\bf (a)} and {\bf (b)}~$g_1$ decomposition at $Q^2=1.01$ and 4.16~GeV$^2$;
{\bf (c)} and {\bf (d)}~$g_2$ decomposition at $Q^2=1.01$ and 4.16~GeV$^2$; 
{\bf (e)} and {\bf (f)}~$F_2$ decomposition at $Q^2=1.01$ and 4.16~GeV$^2$.
Here, the kinematic factors are
$\alpha=M^2/(1+Q^2/\nu^2)$, 
$\beta=\alpha\sqrt{2Q^2}/\nu$, 
$\gamma=\alpha\nu\sqrt{2/Q^2}$, and 
$\delta=M\nu/(1+\nu^2/Q^2)$.}
\label{F:gpm0Dec}
\end{figure*}

To further understand the origins of the behaviors observed in the structure functions in Figs.~\ref{F:g1sing}--\ref{F:h1232sing}, in Fig.~\ref{F:gpm0Dec} we show the decomposition of the computed resonant contributions to $g_1$, $g_2$, and $F_2$ into the individual terms given in Eqs.~(\ref{Eq:coherent}).
This amounts to the description in terms of nucleon resonance electroexcitations by transversely polarized virtual photons of helicities $+1$ (for $G_+$) and $-1$ (for $G_-$), by longitudinally polarized virtual photons ($G_0$) in the virtual photon--target proton center of mass frame, and the mixing term $G_+ G_0$, which describes the interference between electroexcitations by longitudinal and transversely polarized virtual photons. 
One can see that, especially for $F_2$, the evolution of the structure functions with $W$ and $Q^2$ in the second and third resonance regions is driven by the transverse electrocoupling $A_{1/2}$ (or $G_+$), while the first resonance region is dominated by the spin-flip electrocoupling $A_{3/2}$ (or $G_-$) of the $\Delta(1232)\, 3/2^+$.
This correlates with the behavior seen in the helicity structure functions $H_{1/2}$ and $H_{3/2}$ in Fig.~\ref{F:h1232sing}. \\

%%%%%%%%%%%%%%%%%%%%%%%%%%%%%%%%%%%%%%%%%%%%%%%%%%%%%%%%%%%%%%%%%%%%%%%%%%%%
\section{Helicity dependence of quark-hadron duality}
\label{sec:duality}

\begin{figure*}[th]
\includegraphics[width=0.83\textwidth]{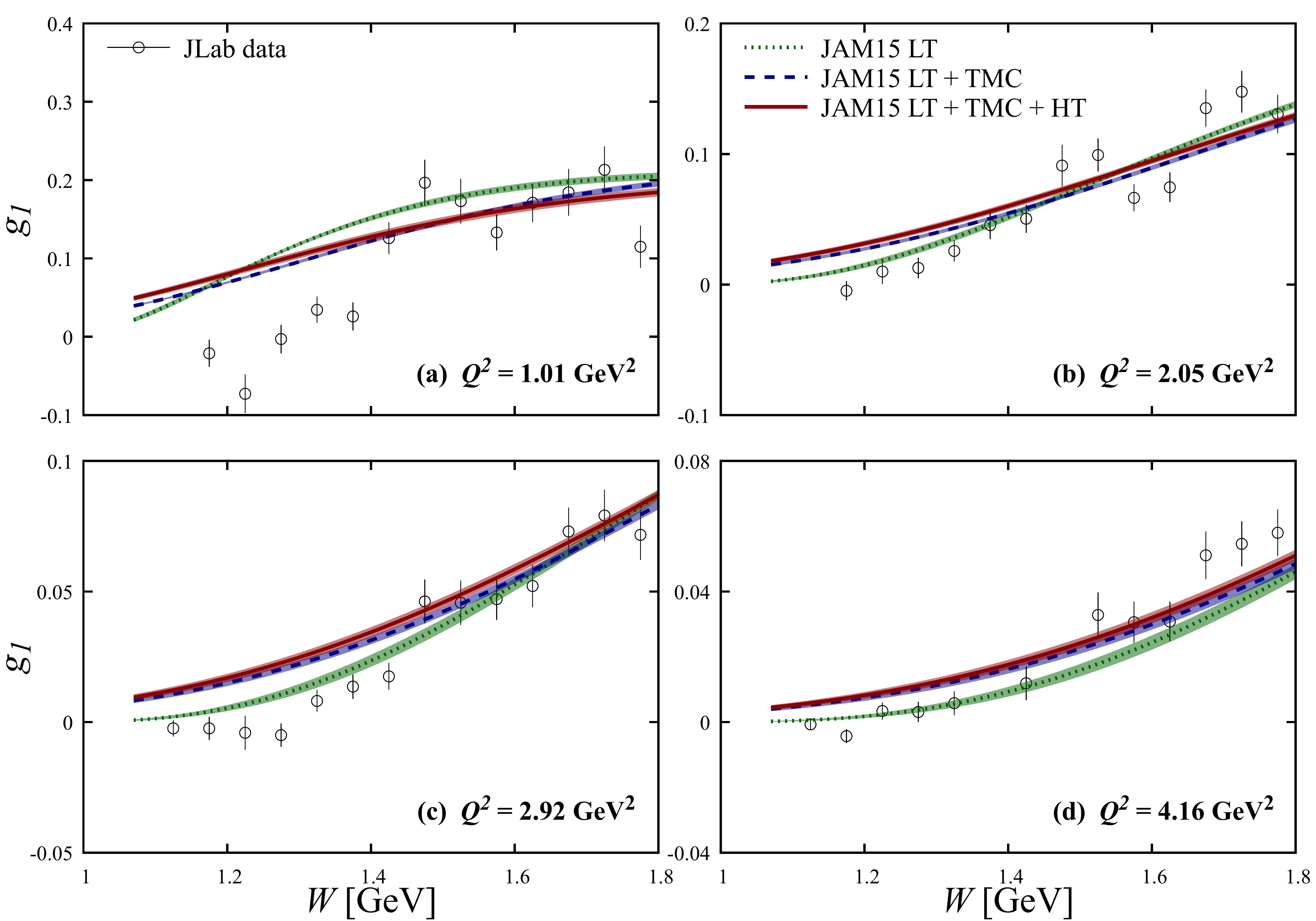}
\caption{Comparison between $g_1$ structure function data from CLAS~\cite{CLAS:SFDB} (open black circles) and $g_1$ computed from PDF-based parametrizations fitted to higher-$W$ data and extrapolated to the resonance region, versus $W$ at fixed values of $Q^2$:
{\bf (a)}~$Q^2=1.01$~GeV$^2$, 
{\bf (b)}~$Q^2=2.05$~GeV$^2$, 
{\bf (c)}~$Q^2=2.92$~GeV$^2$, 
{\bf (d)}~$Q^2=4.16$~GeV$^2$.
The PDF-based calculations are derived from the JAM15 global QCD analysis~\cite{Sato:2016tuz} using leading twist (LT) contributions only (dotted lines), including target mass corrections (TMC) (dashed lines), and TMC and higher twist (HT) contributions (solid lines). The experimental errors include statistical and systematic uncertainties, while the colored bands around the curves represent uncertainties in the high energy scattering data to which the PDFs are fitted using the Monte Carlo technique described in Ref.~\cite{Sato:2016tuz}.}
\label{F:g1pdf}
\end{figure*}

To explore the details of the duality between the low-$W$ resonance region and the high-$W$ deep-inelastic scattering region, in Fig.~\ref{F:g1pdf} we compare the $g_1$ structure function data from CLAS~\cite{CLAS:SFDB} with $g_1$ computed from PDF parametrizations fitted to the DIS data and extrapolated to the resonance region. 
We compare the curves corresponding to the PDF-based parametrization from the JAM15 global QCD analysis \cite{Sato:2016tuz} with those that also account for target mass corrections (TMCs), as well as with results that include both TMCs and higher twist effects.
The inclusion of TMCs and higher twists is expected to give the most reliable results when considering the transition to lower energies and assessing the degree of validity of duality between the resonance and deep-inelastic contributions. 
The structure functions are shown as a function of $W$ at fixed values of $Q^2$ between $Q^2 \approx 1$~GeV$^2$ and 4~GeV$^2$.

With the exception of the first resonance region that is dominated by the $\Delta(1232)\,3/2^+$ resonance, there is a clear similarity in the magnitude of the resonance and extrapolated deep-inelastic contributions, even down to low $Q^2$ values $\sim 1$~GeV$^2$.
This was qualitatively observed already in the early spin structure function measurements~\cite{Baum:1980mh, Melnitchouk:2005zr}, but is confirmed more dramatically with the recent high-precision data.
Unlike the extrapolated deep-inelastic functions, the $\Delta(1232)\,3/2^+$ resonance contribution to $g_1$ remains negative, especially at low $Q^2$.
Once more, this is due to the larger size of the $A_{3/2}$ electrocoupling for this resonance.
In practice, since the PDF-based fit extrapolated from large-$W$ remains positive in this region, there will be a greater mismatch between the parton-level and hadron-level results, and one would therefore expect quark-hadron duality to set in at higher $Q^2$ values, where the $\Delta(1232)\,3/2^+$ contribution is relatively smaller, for spin-dependent structure functions than for the corresponding unpolarized scattering observables~\cite{Blin:2021twt}.

On the other hand, the resonant contributions to $g_1$ computed from exclusive meson electroproduction data~\cite{Mokeev:2022xfo, Carman:2020qmb} are revealed to be substantial in the resonance region at $Q^2 \lesssim 4$~GeV$^2$, and in some cases even dominant.
It has been hypothesized (see Ref.~\cite{Melnitchouk:2005zr} and references therein) that if one could quantify the size of the duality violations by accounting for the resonant contributions evaluated with nucleon resonance electrocouplings, it may then be possible to extend our knowledge of spin-dependent nucleon PDFs to regions at larger $x$ values than is currently possible in global QCD analyses.
Empirical input on the electrocouplings may therefore mitigate the systematic uncertainties stemming from the separation between resonant and nonresonant contributions.

Of course, the degree to which quark-hadron duality can hold locally is naturally limited by the existence of a strong $W$ dependence associated with the resonance structures, so that duality can in practice never hold at all $x$ values.
Any analysis of duality must therefore involve some averaging over $W$ in order to establish quantitative measures of its validity or violation.
On the other hand, a part of the uncertainty in the averaging procedure is precisely the choice of bounds on the $W$ regions over which to average, or on the specific averaging procedures.
To minimize the dependence on the assumptions about the averaging procedures when comparing between the behavior of the averaged resonance peaks with the smooth behavior of the functions extrapolated from the DIS region to low $W$, one can consider so-called ``truncated'' moments, whose evaluation is entirely data driven~\cite{Forte:1998nw, Forte:2000wh, Piccione:2001vf, Kotlorz:2006dj, Psaker:2008ju, Kotlorz:2016icu}.
One of the advantages of this approach is that the $Q^2$ dependence of the leading twist truncated moments is governed by the same $Q^2$ evolution equations that apply for the PDFs themselves.

We define the lowest truncated moments of the $g_1$ and $g_2$ structure functions in an interval $\Delta \xb \equiv x_{\rm max} - x_{\rm min}$ at a fixed $Q^2$ value as
\begin{align}
\label{moments}
%    \Gamma_{1,2}(x_{\rm min}, x_{\rm max}; Q^2)
    \Gamma_{1,2}(\Delta x; Q^2)
%    = \int_{x_{\rm min}}^{x_{\rm max}} \mathrm{d}\xb\, g_{1,2}(x,Q^2).
    = \int_{\Delta x} \mathrm{d}\xb\, g_{1,2}(x,Q^2).
\end{align}
Experimentally, the truncated inelastic moments can be evaluated in regions between the pion production threshold, \mbox{$W_\pi = M + m_\pi$}, where $m_\pi$ is the pion mass, and the maximal value of $W \approx 1.75$~GeV where $\gamma^* p N^*$ electrocouplings are currently available~\cite{Carman:2020qmb}.
Typically, one considers the definite-$W$ intervals
(i)   $W = [W_\pi, 1.38]$\,GeV,
(ii)  $W = [1.38, 1.58]$\,GeV, and
(iii) $W = [1.58, 1.75]$\,GeV, 
corresponding to the first, second, and third resonance regions, respectively.
The integration limits in Eq.~(\ref{moments}) %$[x_\text{min},x_\text{max}]$
corresponding to these $W$ intervals are of course $Q^2$ dependent.
In practice, the truncated moments of the interpolated experimental data on $g_{1,2}$~\cite{CLAS:SFDB} can be evaluated as discrete sums,
\begin{align}
g_{1,2}^{\rm exp} = \sum_i\, \mathrm{d}x_i\, g_{1,2}^i(x_i,Q^2),
\end{align}
where $i$ runs over all the bins of size $\mathrm{d}x_i$ for which $x_{\rm min} \leq x_i \leq x_{\rm max}$, and $g_{1,2}^i$ is the value of the structure function in that bin.

\begin{figure*}[th]
\includegraphics[width=0.84\textwidth]{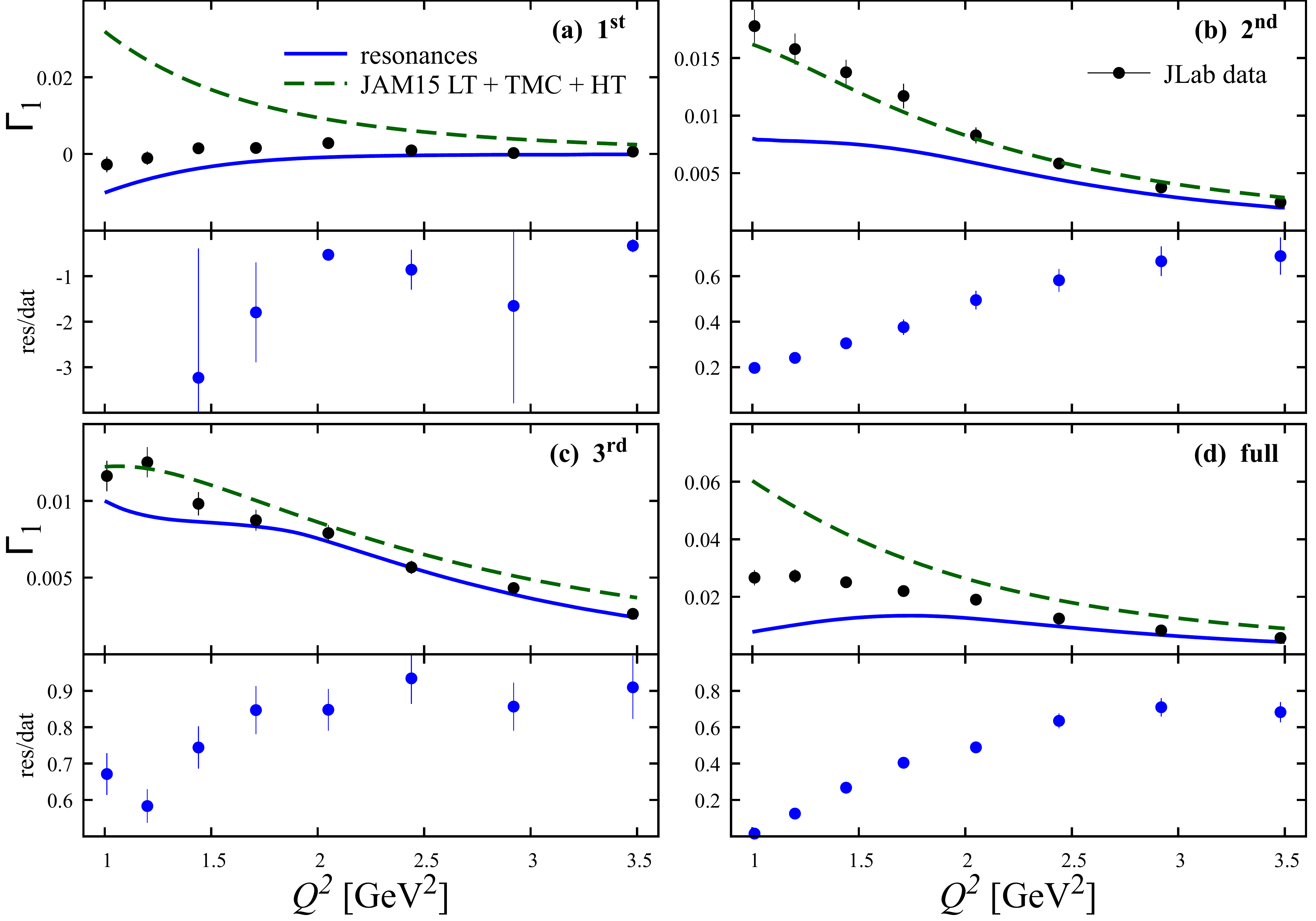}
\caption{Truncated moments $\Gamma_1(\Delta x;Q^2)$ of the $g_1$ structure function versus $Q^2$ for
{\bf (a)} first, or $\Delta(1232)\,3/2^+$, 
{\bf (b)} second, and
{\bf (c)} third resonance regions, and 
{\bf (d)} full resonance region from the pion threshold to $W=1.75$~GeV. 
The moments from the experimental results~\cite{CLAS:SFDB} (black circles) are compared with the resonant contributions (blue lines) and the extrapolated functions from the JAM15 fits to the DIS region \cite{Sato:2016tuz} (green dashed lines). In the lower part of each of the four panels, we also show the ratios of the resonant contributions to the total data, with data uncertainties propagated.}
\label{F:g1trunc}
\end{figure*}

\begin{figure*}[th]
\includegraphics[width=0.84\textwidth]{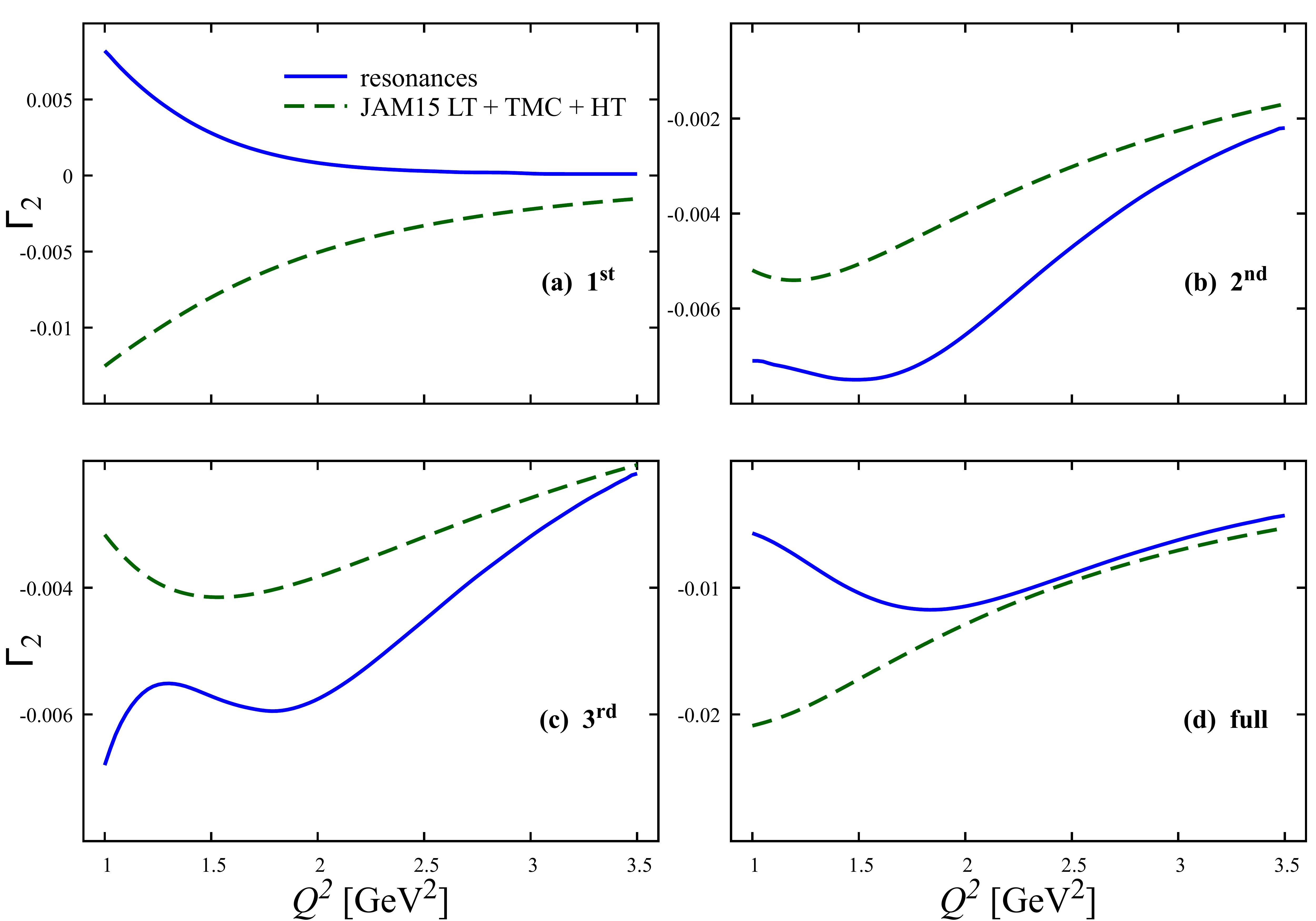}
\caption{As in Fig.~\ref{F:g1trunc}, but for the truncated moments $\Gamma_2(\Delta x;Q^2)$ of the $g_2$ structure function.}
\label{F:g2trunc}
\end{figure*}

The results for the $\Gamma_1$ and $\Gamma_2$ truncated moments are shown in Figs.~\ref{F:g1trunc} and \ref{F:g2trunc}, respectively, for $Q^2$ between 1.0 and 3.5~GeV$^2$.
For $\Gamma_1$ the calculated resonance contributions are also compared with the experimental values obtained from CLAS high-acceptance data, where available.
To quantify the impact of the resonant contributions on the evolution of $\Gamma_1$ with $Q^2$, we evaluate the ratios of the truncated moments of the resonant contributions to those for the full truncated moment in the same $\Delta x$ interval.
These are displayed in the lower part of each panel in Fig.~\ref{F:g1trunc}.

With the exception of the $\Delta(1232)\,3/2^+$, at low $Q^2$ the resonant contributions to $\Gamma_1$ display a clear increase with $Q^2$, before leveling off for $Q^2 \gtrsim 2.5$~GeV$^2$.
For the first resonance region, due to the dominance of the $A_{3/2}$ amplitude for the $\Delta(1232)\,3/2^+$ resonance, the ratio of the resonance contribution to data is negative over most of the $Q^2$ range considered, as may be expected.
In the second and third resonance regions, the resonance contributions to the data are $\sim$ 60\% and 90\%, respectively.
For the entire resonance region up to $W=1.75$~GeV, the overall level of the resonant contribution to the data is also $\sim 60\%$.

Interestingly, the JAM15 global QCD fit \cite{Sato:2016tuz}, extrapolated to the low-$W$ region, is able to describe well the $\Gamma_1$ experimental data in the second and third resonance regions, suggesting that duality violations are not large in these regions.
In contrast, violations of duality are rather significant in the first resonance region, closest to threshold, where both the magnitude and sign of the experimental truncated $\Gamma_1$ moment differ from the naive DIS-extrapolated result.
This places clear limits on the extent to which low-$W$ data may be utilized to infer partonic information, even in a quark-hadron duality averaged sense.

For the $\Gamma_2$ truncated moments in Fig.~\ref{F:g2trunc}, since there are no high-acceptance data available for the $g_2$ structure function, we focus our attention on the comparison between the resonant contributions and DIS fit extrapolations. 
Generally, the sign of $\Gamma_2$ is found to be opposite to that of $\Gamma_1$ seen in Fig.~\ref{F:g1trunc}.
There is a large difference, in both sign and magnitude, between the resonance contributions and the DIS-extrapolated results in the first resonance region.
On the other hand, one finds a convergence of the resonance and DIS-extrapolated results  at larger $Q^2$ values in the second and third regions.
Furthermore, there is a remarkable similarity between the two in the full resonance region for $Q^2 \gtrsim 2$~GeV$^2$, and it will be interesting to confront these predictions with future measurements of $\Gamma_2$.
In fact, since in the computation of the $g_2$ structure function the nonperturbative quark-gluon effects enter at the same order as the leading twist contributions, measurements of $g_2$ may provide the most direct access to quark-hadron duality studies~\cite{Melnitchouk:2005zr}.
% This strongly motivates further studies of the inclusive structure functions at ongoing measurements with detectors of nearly $4\pi$ acceptance, such as the CLAS12 endeavors at JLab and in future experiments after a CEBAF energy increase.

%%%%%%%%%%%%%%%%%%%%%%%%%%%%%%%%%%%%%%%%%%%%%%%%%%%%%%%%%%%%%%%%%%%%%%%%%%%%
\section{Summary and outlook}
\label{sec:outlook}

In this work we have studied the role of nucleon resonances in spin-dependent observables in inclusive electron scattering from proton targets.
Using empirical input for electroexcitation amplitudes extracted from CLAS data, we evaluated the coherent sum of resonances contributing to the spin-dependent $g_1$ and $g_2$ structure functions, computing their $W$ and $Q^2$ dependence for the resonances in the mass range up to $W=1.75$~GeV where the experimental results on resonance electrocouplings are available.
Detailed comparisons between the resonance contributions and inclusive scattering data allowed us to quantify the spin dependence of quark-hadron duality in the transition between the low-energy regime, driven by resonance excitations, and the DIS region dominated by scattering from partons.

In all the spin-dependent observables considered, the behavior of most of the resonances is driven by the larger values of the  $A_{1/2}$ electrocouplings compared with the $S_{1/2}$ and $A_{3/2}$ electroexcitation amplitudes.
The exceptions are the $\Delta(1232)\,3/2^+$ and the $N(1720)\,3/2^+$ states, where $|A_{3/2}| > |A_{1/2}|$ over the entire range of $Q^2$ considered.
Since the $\Delta(1232)\,3/2^+$ almost saturates the resonance contributions in the first resonance region, a sign flip is seen in the $W$ dependence of the resonance contributions to both $g_1$ and $g_2$ between the first and the second resonance peaks.
This behavior has previously been observed in the inclusive $g_1$ data, and is also expected to hold for $g_2$ if the Burkhardt-Cottingham sum rule~\cite{Burkhardt:1970ti} is valid.
Our findings provide evidence for the first time that this behavior is indeed accounted for by the resonance contributions.

Confirming that this result also holds for the $g_2$ structure function gives clear motivation for future large-acceptance measurements of this observable.
To date, for each $Q^2$ bin, $g_2$ data in the resonance region have only been available for a narrow angular acceptance, washing out the resonance peaks that would otherwise be visible in the $W$ dependence, as was confirmed in our computations.

While in the second resonance region the shape of the $W$ dependence of all the observables can be traced back to, or even saturated by, the $N(1520)\,3/2^-$ and $N(1535)\,1/2^-$ states, in the third resonance region we find different resonance peaks and tails to be dominant, depending on the observable.
This underlines the importance of a systematic study of all resonance electrocouplings in ongoing and future experiments, such as those with CLAS12 at Jefferson Lab.

Our analysis was able to confirm a duality between the $g_1$ resonance region data and parametrizations of high-energy data extrapolated down to low energies, especially in the second and third resonance regions.
This becomes more apparent when quantifying the duality in the form of truncated moments of structure functions, which provide a more robust method of averaging over the nontrivial peaks in the $W$ dependence of the resonance region data.
It is also evident from our studies that, particularly for the first resonance region, where the structure functions display a sign flip, there are limits to the extent to which duality can be local.
This is not surprising given that the $\Delta(1232)\,3/2^+$ is the dominant resonance in this region, with minimal overlap from other states, and furthest from the region that is used to constrain PDF fits.

Our findings provide motivation for further exploration of the possibility of lowering the $W$ cuts on inclusive scattering data below the typical $W \gtrsim 2$~GeV cut used in global QCD analyses.
Moreover, future data in the transition region between the resonance and DIS regimes will provide further insight into the connection between the physics of quark-gluon dynamics which underlies the generation of the ground and excited states of the nucleon.
Experiments measuring the $g_2$ structure function with high-acceptance detectors, such as CLAS12, may be a promising avenue for duality studies, due to the emergence of nonperturbative quark-gluon dynamics that enter at the same order as the leading twist terms.

Ultimately, the goal is to have a theoretically well founded, as well as data driven,  description of the transition from the perturbative regime of quarks and gluons to the low-$W$ and $Q^2$ region that is most efficiently described in terms of hadron degrees of freedom.
Isolating the description of the resonance contributions is therefore a benchmark that mitigates the systematic uncertainties stemming from the method of separating resonant from nonresonant contributions in a smooth transition across energies and photon virtualities.

The codes to generate the results presented in this article are available online~\cite{github:ANHB}.

%%%%%%%%%%%%%%%%%%%%%%%%%%%%%%%%%%%%%%%%%%%%%%%%%%%%%%%%%%%%%%%%%%%%%%%%%%%%
\begin{acknowledgements}
We thank C.~Cocuzza and N.~Sato for providing the NLO structure function calculation code used in our calculations, and O.~Rondon, W.~Armstrong, and the CLAS Physics database manager V.~V.~Chesnokov for the experimental results on polarized asymmetries. We also thank S.~Kuhn, V.~Lagerquist, and P.~Pandey for useful discussions.
This work was supported by the U.S. Department of Energy Contract No.\ DE-AC05-06OR23177, under which Jefferson Science Associates, LLC operates Jefferson Lab, and by the Deutsche Forschungsgemeinschaft (DFG) through the Research Unit FOR 2926 (Project No.\ 409651613).
\end{acknowledgements}

% \newpage
%%%%%%%%%%%%%%%%%%%%%%%%%%%%%%%%%%%%%%%%%%%%%%%%%%%%%%%%%%%%%%%%%%%%%%%%%%%%%
\bibliographystyle{apsrev4-1-jpac}
\bibliography{cite}{}

\end{document}